\documentclass[aps,pre,twocolumn,groupedaddress,amsmath]{revtex4}

\usepackage{graphicx}

\allowdisplaybreaks

\begin{document}

\title{Chiral and Antichiral Order in Bent-Core Liquid Crystals}

\author{Jonathan V. Selinger}
\affiliation{Center for Bio/Molecular Science and Engineering,
Naval Research Laboratory, Code 6900,\\
4555 Overlook Avenue, SW, Washington, DC 20375}

\date{March 18, 2003}

\begin{abstract}
Recent experiments have found a bent-core liquid crystal in which the layer
chirality alternates from layer to layer, giving a racemic or ``antichiral''
material, even though the molecules are uniformly chiral.  To explain this
effect, we map the liquid crystal onto an Ising model, analogous to a model for
chiral order in polymers.  We calculate the phase diagram for this model, and
show that it has a second-order phase transition between antichiral order and
homogeneous chiral order.  We discuss how this transition can be studied by
further chemical synthesis or by doping experiments.
\end{abstract}

\pacs{61.30.Dk, 64.70.Md}

\maketitle

Many types of soft condensed matter exhibit chiral order on the macroscopic
length scale.  In some materials, macroscopic chiral order is a direct
consequence of molecular chirality.  For example, in cholesteric liquid
crystals, the packing of neighboring chiral molecules leads to a large-scale
helical twist of the molecular orientation~\cite{degennes93,harris99,kamien01}.
However, in other materials, the relationship between molecular chirality and
macroscopic chiral order is more complex.  Some liquid crystals develop chiral
order through a spontaneous symmetry-breaking transition, although they are
composed of achiral molecules~\cite{selinger93,pang94}.  Likewise, some polymers
amplify slight chiral perturbations to give highly ordered helical
chains~\cite{green99}.  In recent experiments, Nakata \emph{et al.}\ have
found a novel relationship between molecular chirality and macroscopic chiral
order in a bent-core liquid crystal~\cite{nakata01}.  This material forms a
smectic phase with layers of alternating right- and left-handed chirality, which
we call ``antichiral'' order, in spite of the fact that it is composed of
uniformly chiral molecules, i.e.\ 100\% of a single enantiomer.  In this paper,
we develop a theory to explain how antichiral order can occur in systems of
chiral molecules.  We predict a phase diagram with a second-order transition
between antichiral order and homogeneous chiral order, and discuss how this
phase diagram can be explored in further experiments on bent-core liquid
crystals.

\begin{figure}[b]
\includegraphics[width=3.375in,clip]{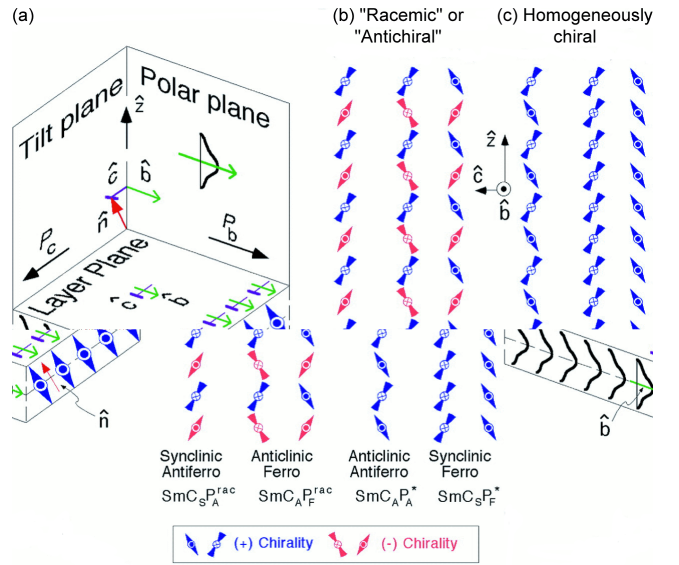}%
\caption{(a)~Geometry of a smectic layer in a bent-core liquid crystal, showing
the tilt order and polar order.  (b)~``Racemic'' or ``antichiral'' order from
layer to layer.  (c)~Homogeneous chiral order from layer to layer.  Adapted from
Ref.~\protect\cite{link97}.}
\end{figure}

Bent-core liquid crystals have been studied extensively since the initial
investigation by Niori \emph{et al.}~\cite{niori96}.  The main reason for this
wide interest is that bent-core liquid crystals can develop spontaneous chiral
order within the layers of a smectic phase. The geometry of this ordering, as
determined by Link \emph{et al.}~\cite{link97}, is shown in Fig.~1(a).  Within
each smectic layer, the molecules can have two types of orientational order:
tilt order (represented by a tilt of the molecular director $\mathbf{\hat{n}}$
with respect to the layer normal $\mathbf{\hat{z}}$) and polar order
(represented by ordering of the molecular polarization vector
$\mathbf{\hat{b}}$).  The combination of tilt order and polar order gives the
layer a chiral structure.  Geometrically, we can see that the structure
characterized by $\mathbf{\hat{n}}$ and $\mathbf{\hat{b}}$ is the mirror image
of the structure characterized by $\mathbf{\hat{n}}$ and $-\mathbf{\hat{b}}$.
Mathematically, we can represent the magnitude and sign of the chiral order by
the order parameter
$\chi=\langle2[(\mathbf{\hat{z}}\times\mathbf{\hat{n}})\cdot\mathbf{\hat{b}}]
(\mathbf{\hat{z}}\cdot\mathbf{\hat{n}})\rangle$~\cite{xu01}.  The mirror-image
structure has the opposite value of $\chi$.

Bent-core liquid crystals can have different types of order from layer to layer.
As shown in Fig.~1(b-c), adjacent smectic layers can be synclinic (same tilt) or
anticlinic (opposite tilt).  Likewise, adjacent layers can be ferroelectric
(same $\mathbf{\hat{b}}$) or antiferroelectric (opposite $\mathbf{\hat{b}}$).
If layers are synclinic and ferroelectric, or if they are anticlinic and
antiferroelectric, then they have the same layer chirality $\chi$.  In this
case, the bulk liquid crystal is homogeneously chiral.  By contrast, if layers
are synclinic and antiferroelectric, or if they are anticlinic and
ferroelectric, then they have the opposite layer chirality $\chi$.  In that
case, the bulk liquid crystal has an alternating chiral structure.  This
alternating structure has been called a ``racemic'' structure by Link
\emph{et al.}~\cite{link97}.  However, we propose the more specific term
``antichiral'' to emphasize that there is a rigid alternation between right- and
left-handed layers, rather than just a 50/50 statistical distribution of right-
and left-handed domains.  The term ``antichiral'' has been used to describe the
packing of right- and left-handed helices in crystals of
polypropylene~\cite{lotz96}, and our usage is consistent with that literature.

If a bent-core liquid crystal is composed of achiral molecules, the right- and
left-handed layer structures are exact mirror images of each other, and hence
they have the same free energy.  For that reason, they should occur equally
often.  By contrast, if a bent-core liquid crystal is composed of chiral
molecules, the molecular chirality breaks the symmetry between right- and
left-handed layer chirality.  In other words, the right- and left-handed layers
become diastereomers rather than enantiomers.  Hence, they do not have the same
free energy, so one structure is energetically preferred compared to the other.
Indeed, adding a low concentration of a chiral dopant to an achiral bent-core
liquid crystal converts most of the liquid crystal to a homogeneously chiral
state of a single handedness~\cite{link97}.  However, Nakata \emph{et al.}\ made
a surprising discovery:  In experiments on well-aligned domains of the chiral
bent-core liquid crystal 8OPIMB6*, they found a smectic phase that was both
anticlinic and ferroelectric~\cite{nakata01}.  This result was consistent with
earlier studies of unaligned cells of the same material~\cite{gorecka00}.  This
combination of tilt and polar order implies that the smectic layers are
antichiral.  Hence, the phase has alternating right- and left-handed layers,
even though the chirality of the molecules must favor one sense of the layer
chirality.

The experiment of Nakata \emph{et al.}\ leads to an important theoretical issue:
What is the relationship between molecular chirality and macroscopic chiral
order in a bent-core liquid crystal?  In particular, how can the macroscopic
order be antichiral when the molecules are uniformly chiral?  The macroscopic
chiral order should respond in some way to the molecular chirality.  We would
like to develop a theory for this relationship.

To address this issue, we map the system of bent-core liquid crystals onto the
antiferromagnetic Ising model in a uniform magnetic field~\cite{lavis99}.  The
Hamiltonian is
\begin{equation}
H=+J\sum_i\sigma_i\sigma_{i+1}-h\sum_i\sigma_i .
\end{equation}
In this mapping, the Ising spin $\sigma_i$ represents the chiral order
$2[(\mathbf{\hat{z}}\times\mathbf{\hat{n}})\cdot\mathbf{\hat{b}}]
(\mathbf{\hat{z}}\cdot\mathbf{\hat{n}})$ of layer $i$.  Hence,
$\langle\sigma_i\rangle=+1$ corresponds to an ideal right-handed layer and
$\langle\sigma_i\rangle=-1$ to an ideal left-handed layer.  A fractional value
of $\langle\sigma_i\rangle$ corresponds either to reduced uniform chiral order
in the layer (e.g.\ a reduced value of the tilt) or to an average over
coexisting regions of different chiral order.  The Ising field $h$ represents
the uniform chirality of the molecules, which favors one sense of the layer
chirality.  The Ising exchange constant $J$ represents the interaction between
adjacent layers.  This constant is positive, so that the layer interfaces tend
to be antichiral (e.g.\ anticlinic and ferroelectric).  This preference for
antichirality presumably arises from steric interactions between molecular tails
at the layer interfaces.

There is a competition between the two terms in the Hamiltonian:  the $h$ term
favors homogeneous chiral order, with the handedness favored by the sign of $h$,
while the $J$ term favors an alternation from layer to layer.  Hence, we need to
solve the model to determine whether the layers are homogeneous or alternating
for particular values of $h$ and $J$.

\begin{figure}
\includegraphics[width=3.375in,clip]{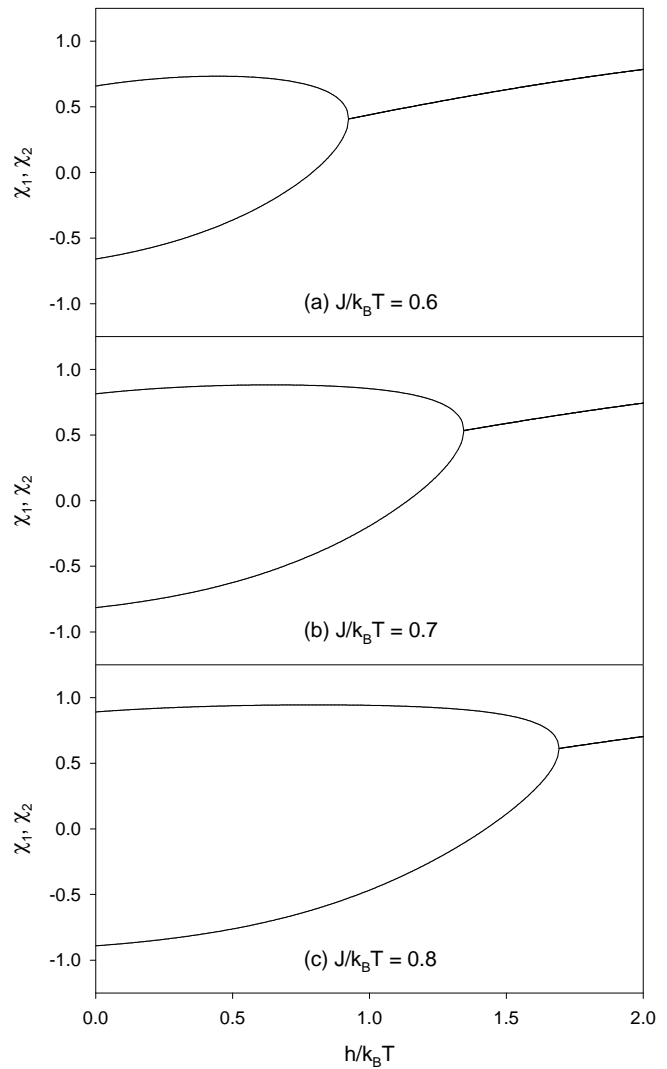}%
\caption{Numerical solution for the chiral order parameters of the odd and even
layers, $\chi_1$ and $\chi_2$, as functions of the molecular chirality $h$, for
three values of the interlayer interaction $J$.  (a)~$J/k_B T=0.6$.
(b)~$J/k_B T=0.7$.  (c)~$J/k_B T=0.8$.  There is a second-order transition
between antichiral order and homogeneous chiral order at a critical value of
$h$, which increases as a function of $J$.}
\end{figure}

To solve this model, we use mean-field theory.  We suppose that each layer
responds to the mean field due to the adjacent layers at the temperature $T$.
We look for a solution with alternating layers
\begin{equation}
\langle{\sigma_i}\rangle=\left\{\begin{array}{ll}
\chi_1 & \mbox{if $i$ is odd,}\\
\chi_2 & \mbox{if $i$ is even.}
\end{array}\right.
\end{equation}
In that case, the mean-field equations become
\begin{subequations}
\label{meanfield}
\begin{eqnarray}
\chi_1&=&\tanh\left(\frac{-2J\chi_2+h}{k_B T}\right),\\
\chi_2&=&\tanh\left(\frac{-2J\chi_1+h}{k_B T}\right).
\end{eqnarray}
\end{subequations}
These equations must be solved self-consistently to obtain $\chi_1$ and $\chi_2$
as functions of $h/k_B T$ and $J/k_B T$.

The numerical solution of the mean-field equations is shown in Fig.~2.  When the
molecular chirality $h=0$, the layer order parameters $\chi_1$ and $\chi_2$ have
equal and opposite values, indicating that odd and even layers are mirror images
of each other.  In that case, the system has perfect antichiral order.  As $h$
increases, the positive order parameter $\chi_1$ becomes slightly larger, and
the negative order parameter $\chi_2$ becomes less negative.  Still, there is an
alternation between odd and even layers with right- and left-handed chiral
order, and hence the system is still antichiral.  This alternating layer
structure persists up to a critical value of $h$, where there is a second-order
transition to homogeneously chiral layers, with the same value of $\chi$.
Beyond that transition, the chiral order parameter of the layers increases
smoothly as a function of $h$.

\begin{figure}
\includegraphics[width=3.375in,clip]{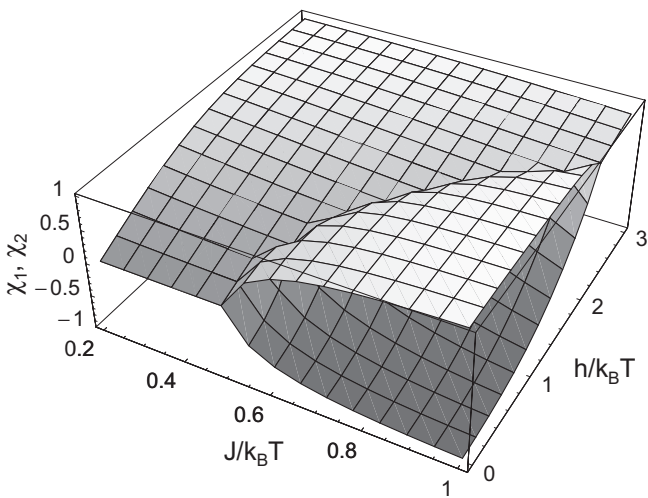}%
\caption{Three-dimensional plot of the layer order parameters $\chi_1$ and
$\chi_2$ as functions of $h/k_B T$ and $J/k_B T$.  The transition between
alternating layers and homogeneously chiral layers can be induced by varying the
molecular chirality $h$, the interlayer interaction $J$, or the temperature
$T$.}
\end{figure}

A three-dimensional plot of $\chi_1$ and $\chi_2$ as functions of $h/k_B T$ and
$J/k_B T$ is presented in Fig.~3.  This plot shows that the antichiral state is
favored for large $J$ and small $h$, while the homogeneously chiral state is
favored for large $h$ and small $J$.  Moreover, the plot shows that the
second-order transition can be driven by varying the molecular chirality $h$,
the interlayer interaction $J$, or the temperature $T$.  Increasing $J$ breaks
the symmetry between the layers, while increasing $h$ or $T$ restores the
symmetry.  Thus, the transition shown in Fig.~2 for varying $h$ is continuously
connected to the standard Ising antiferromagnetic transition for varying $T$ at
$h=0$.

\begin{figure}
\includegraphics[width=3.375in,clip]{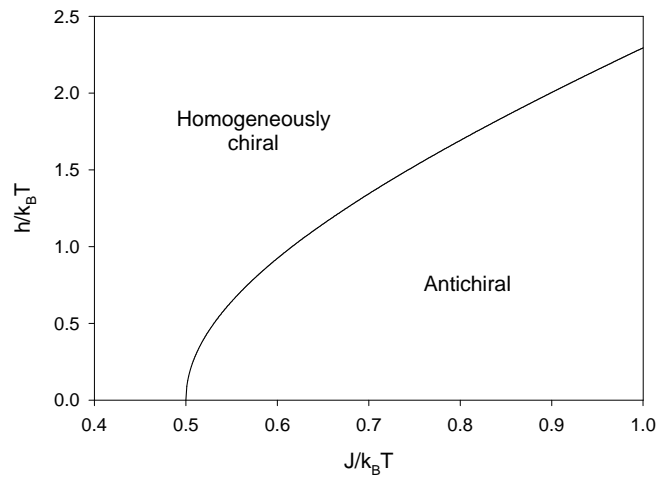}%
\caption{Phase diagram in terms of $h/k_B T$ and $J/k_B T$.  The antichiral
state is favored for high interlayer interaction $J$ and low molecular chirality
$h$, while the homogeneously chiral state is favored for high $h$ and low $J$.}
\end{figure}

In Figs.~2 and~3, we see that the critical value of $h$ for the
chiral-antichiral transition depends on $J$.  As $J$ increases, the alternating
layer structure persists to higher values of $h$.  To find the critical value
$h_c$ as a function of $J$, we look for a solution of the mean-field
equations~(\ref{meanfield}) with the layer order parameters
$\chi_{1,2}=\chi_c\pm\delta\chi$.  As we approach the transition, we have
$\delta\chi\to0$, and hence the mean-field equations require
\begin{subequations}
\begin{eqnarray}
\chi_c&=&\tanh\left(\frac{-2J\chi_c+h}{k_B T}\right),\\
\delta\chi&=&\frac{2J\delta\chi}{k_B T}
\mbox{sech}^2\left(\frac{-2J\chi_c+h}{k_B T}\right).
\end{eqnarray}
\end{subequations}
The solution of these equations is
\begin{subequations}
\begin{eqnarray}
\chi_c&=&\sqrt{1-\frac{k_B T}{2J}},\\
\frac{h_c}{k_B T}&=&\frac{2J}{k_B T}\sqrt{1-\frac{k_B T}{2J}}
+\tanh^{-1}\sqrt{1-\frac{k_B T}{2J}}.
\label{phaseboundary}
\end{eqnarray}
\end{subequations}
Equation~({\ref{phaseboundary}) gives the phase boundary between the antichiral
state and the homogeneously chiral state.  This boundary is plotted in the phase
diagram of Fig.~4.

So far, the experiments of Nakata \emph{et al.}~\cite{nakata01} have studied the
antichiral state in the single chiral bent-core liquid crystal 8OPIMB6*.  In
this compound, the chiral centers are located near the ends of the flexible
hydrocarbon chains, far from the bent core of the molecule.  For that reason,
the chiral centers should only have a small effect on the relative energies of
the right- and left-handed layer structures.  In other words, for these
molecules, the parameter $h$ should be low.  By contrast, the parameter $J$
arises from the interfacial energy of adjacent smectic layers, which tend to be
anticlinic and ferroelectric, and hence $J$ should be substantial.  The regime
of small $h$ and large $J$ is consistent with the antichiral state in the
theoretical phase diagram.

To map out the rest of the phase diagram, future experiments will need to vary
$h$ and $J$.  The molecular chirality parameter $h$ can be changed either
through chemical synthesis or through doping experiments.  For the synthetic
approach, one would prepare molecules analogous to 8OPIMB6*, but with the chiral
centers closer to the bent core.  By synthesizing a series of molecules with
the chiral centers in different positions, one could gradually increase $h$.
Alternatively, one could dope 8OPIMB6* with varying concentrations of a chiral
dopant---perhaps another bent-core liquid crystal that forms a homogeneously
chiral state.  Increasing the dopant concentration should gradually increase
$h$.  The doping approach should actually increase $h$ in a more quantitatively
controllable way than the synthetic route, as long as the compounds do not phase
separate.  In either case, we would expect to see a transition from the
antichiral state to the homogeneously chiral state as $h$ increases.

Changing the interlayer interaction $J$ will require changes in the layer
interfaces.  In particular, to explore the theoretical phase diagram, one will
need to reduce the energetic preference for anticlinic and ferroelectric packing
of adjacent layers.  Hence, one should partially decouple the layers.  This
might be done by doping 8OPIMB6* with molecules that segregate into the
interstices between the layers (analogous to the nanophase segregation seen in
simulations by Maiti \emph{et al.}~\cite{maiti02}).  The parameter $J$ would
then be a continuously decreasing function of dopant concentration.  We would
expect to see a transition from antichiral to homogeneously chiral as more
dopant is added and the layers become more decoupled.

As a final point, our predictions for bent-core liquid crystals can be compared
with chiral ordering in polymers.  There have been several experimental and
theoretical studies of chiral ordering in polyisocyanates~\cite{green99}.  In
these materials, steric constraints force the polymer main chain to have a
helical structure, which can be either right- or left-handed.  The handedness of
the helix is determined by slight chiral perturbations, such as chiral centers
in the monomers.  For example, copolymers of right- and left-handed units have
been called ``majority-rule'' copolymers, because they follow whichever chiral
units are in the majority.  Likewise, copolymers of chiral and achiral units
have been called ``sergeants-and-soldiers'' copolymers, because the achiral
``soldiers'' follow the helical sense selected by the chiral ``sergeants.''
This chiral ordering has been explained theoretically by a mapping onto the
Ising model~\cite{selinger96,selinger97}.

In both polyisocyanates and bent-core liquid crystals, the chiral order of the
large-scale structure is a complex collective phenomenon.  In both cases, a
theory based on the Ising model can explain the  nonlinear relationship between
molecular chirality and macroscopic chiral order.  However, there is one
important difference between polyisocyanates and bent-core liquid crystals.  In
polyisocyanates, the interaction between neighboring units along the polymer
favors \emph{uniform} helicity.  For that reason, polyisocyanates respond
sensitively to slight chiral perturbations.  By contrast, in the bent-core
liquid crystal 8OPIMB6*, the interaction between adjacent smectic layers favors
\emph{alternating} layer chirality.  Hence, these materials resist uniform
chiral order, and respond insensitively to molecular chirality.  For that
reason, they might be called a ``sergeants-and-students'' system.

In conclusion, we have developed a theory for antichiral order in bent-core
liquid crystals.  This theory explains the surprising experimental discovery
that antichiral order can occur even in systems of uniformly chiral molecules.
It predicts a phase diagram for the antichiral state and the homogeneously
chiral state, which can be explored in future experiments.

We thank D.~R. Link and M. Nakata for helpful discussions.  This work was
supported by the Office of Naval Research and the Naval Research Laboratory.

\end{document}